\documentclass[twocolumn,showpacs,preprintnumbers,amsmath,amssymb]{revtex4-1}

\usepackage{graphicx}
\usepackage{dcolumn}
\usepackage{bm}
\usepackage{longtable}
\usepackage{hyperref}
\usepackage{lipsum}
\usepackage{braket}
\usepackage{physics}
\usepackage{setspace}
\usepackage{color}
\usepackage{fancyhdr}
\usepackage{titlesec}
\usepackage{float}
\usepackage{upgreek}
\usepackage{array}
\usepackage{epstopdf}

\usepackage[utf8]{inputenc}

\usepackage{wrapfig}
\usepackage{amsmath,amsthm,amssymb}
\usepackage[table,xcdraw]{xcolor}

\titleformat*{\section}{\LARGE\bfseries}
\titleformat*{\subsection}{\Large\bfseries}
\titleformat*{\subsubsection}{\large\bfseries}

\begin{document}
\preprint{APS/123-QED}
\title{Geometrical dependence of domain wall propagation and nucleation fields in magnetic domain wall sensor devices.}

\author{B. Borie}%

\affiliation{Institut f\"ur Physik, Johannes Gutenberg-Universit\"at Mainz, Staudinger Weg 7, 55128 Mainz, Germany}%
\affiliation{Sensitec GmbH, Hechtsheimer Str. 2, Mainz D-55131, Germany}%

\author{A. Kehlberger}%
\affiliation{Sensitec GmbH, Hechtsheimer Str. 2, Mainz D-55131, Germany}%

\author{J. Wahrhusen}%
\affiliation{Sensitec GmbH, Hechtsheimer Str. 2, Mainz D-55131, Germany}%

\author{H. Grimm}%
\affiliation{Sensitec GmbH, Hechtsheimer Str. 2, Mainz D-55131, Germany}%

\author{M. Kl\"aui}%
\affiliation{Institut f\"ur Physik, Johannes Gutenberg-Universit\"at Mainz, Staudinger Weg 7, 55128 Mainz, Germany}%
\email{Contact: klaeui@uni-mainz.de}

\date{\today}

\begin{abstract}
We study the key domain wall properties in segmented nanowires loop-based structures used in domain wall based sensors. The two reasons for device failure, namely the distribution of domain wall propagation field (depinning) and the nucleation field are determined with Magneto-Optical Kerr Effect (MOKE) and Giant Magnetoresistance (GMR) measurements for thousands of elements to obtain significant statistics.
Single layers of Ni$_{81}$Fe$_{19}$, a complete GMR stack with Co$_{90}$Fe$_{10}$/Ni$_{81}$Fe$_{19}$ as a free layer and a single layer of Co$_{90}$Fe$_{10}$ are deposited and industrially patterned to determine the influence of the shape anisotropy, the magneto-crystalline anisotropy, and the fabrication processes. We show that the propagation field is little influenced by the geometry but significantly by material parameters. Simulations for a realistic wire shape yield a curling mode type of the magnetization configuration close to the nucleation field. Nonetheless, we find that the domain wall nucleation fields can be described by a typical Stoner-Wohlfarth model related to the measured geometrical parameters of the wires and fitted by considering the process parameters. The GMR effect is subsequently measured in a substantial number of devices (3000), in order to accurately gauge the variation between devices. This reveals a corrected upper limit to the nucleation fields of the sensors that can be exploited for fast characterization of working elements.

\end{abstract}

\maketitle

\section{Introduction}

Magnetic domain walls in soft thin films nanostructures \cite{MK08} are interesting objects for numerous applications such as logic \cite{DAA05} and memory devices \cite{SSPP08}. Furthermore, several sensor ideas based on domain walls have been developed and reported in the literature \cite{HCL15, MAB11, MD09}.

The potential of magnetic domain wall based sensors is due to their many attractive attributes compared to other technologies. Magnetic domain walls can be stable well above room temperature, making it a potential candidate to store data and be used for non-volatile sensing and they can be displaced rapidly in application's relevant geometries \cite{AB13}. The second point of such technologies relies on the fact that no external power is required to create, stabilize or manipulate the storing element, meaning that a power failure in the system does not affect the functionality of the device. This ensures non-stop sensing even in cases where power is lost. Moreover, the only necessary power is that associated with the injected current that reads the state of the device, which can be applied during only a small part of the total operating time. Finally, the integration of the technology is relatively cheap allowing for an abundance of sensors in the desired system.

The first such device to be implemented \cite{MD09} was designed a few years ago by Novotechnik \cite{N} and is now produced by Sensitec \cite{S}. The purpose of this sensor is to count the number of rotations of a magnetic field. The sensor relies on the Giant Magnetoresistance (GMR) \cite{MNB88, GB89} effect to generate the signal. The information of absolute rotation is stored by the use of a combination of determining the domain wall positions and their number within the device. This simple method provides an elegant and reliable solution, which enables a versatile sensor design as required by the application.
The structure, under a rotating applied magnetic field, nucleates one domain wall every 180$^\circ$ rotation, and hence by counting the number of domain walls, the number of 360$^\circ$ turns is extracted (see Supplemental Material for the details on the functionality of the device). These magnetic domain wall devices exhibit two types of failure events, the pinning of domain walls if a particular propagation field threshold is not reached and the undesired nucleation of domain walls at too high fields. Those two critical fields ought to be as separated as possible to allow an extensive range of operating applied fields and thus of industrial applications. 

In this paper, we study the field operating window by measuring the successful operations and errors occurring in the propagation and the nucleation field for several materials and different geometries. We employ Magneto-Optical Kerr Effect \cite{JK77}  microscopy for the precise investigation of the position of domain walls in the devices under various conditions and measure the GMR effect for high-statistics device characterization to obtain relevant information for the industry. The propagation field is identified as being affected by the edge roughness and the crystalline structure of the wires, while the nucleation field is dependent on the shape of the sample and the processing parameters. Simulations are used to identify the 2-dimensional variations of the magnetization close to the nucleation field and the expected nucleation field value. Furthermore, comprehensive statistics are obtained with the use of an automatized measurement using the GMR effect to compare with the MOKE microscopy results and determine a precise nucleation field for a significant number of devices. 

\section{The investigated systems}
                                                                                                                                                                                                                                                                                                                                                                                                                                                                                                                                                                                                                                 \begingroup
\begin{table}[H]
\centering
\resizebox{\columnwidth}{!}{%
\begin{tabular}{c|c|c|c|}
\cline{2-4}
 & \cellcolor[HTML]{32CB00}\textbf{Ni$_{81}$Fe$_{19}$ (28 nm)} & \cellcolor[HTML]{34CDF9}\textbf{Ni$_{81}$Fe$_{19}$ (32 nm)} & \cellcolor[HTML]{FFC702}\textbf{Co$_{90}$Fe$_{10}$ (17 nm)} \\ \hline
\multicolumn{1}{|c|}{\textbf{Nominal Width (nm)}} & \multicolumn{3}{c|}{\textbf{Average Width (nm)}} \\ \hline
\multicolumn{1}{|c|}{200} & 205$\pm$9 & 240$\pm$15 & 211$\pm$6 \\ \hline
\multicolumn{1}{|c|}{250} & 264$\pm$10 & 288 $\pm$15 & 266$\pm$6 \\ \hline
\multicolumn{1}{|c|}{300} & 328$\pm$6 & 342$\pm$9 & 323$\pm$9 \\ \hline
\multicolumn{1}{|c|}{350} & 372$\pm$18 & 391$\pm$9 & 357$\pm$19 \\ \hline
\end{tabular}
}
\caption{Summary of the measured samples. The average width is determined from the average of 15 width measurements from Scanning Electron Microscopy micrographs of the wires.}
\label{fig0}
\end{table}
\endgroup

The samples were deposited in a magnetron sputtering system employing a seed layer. The investigated systems are two single layers of Ni$_{81}$Fe$_{19}$(28 nm and 32 nm), and one single layer of Co$_{90}$Fe$_{10}$(17 nm) (referenced in the Table \ref{fig0}). All the samples were capped with a 4 nm Ta layer. For the GMR measurement, a complete GMR stack with a free layer of Co$_{90}$Fe$_{10}$(1 nm)/Ni$_{81}$Fe$_{19}$(32 nm) was used. The GMR ratio was measured with a four-point probe technique under an applied magnetic field. A resist was spin-coated on the as-deposited wafers and patterned with photolithography in the shape of the structures (see Fig. \ref{fig1}). After the development of the resist, the material was then etched away by an Ar$^ +$ ion etching process.

To fabricate electrical connections, part of the batch went through a second lithography followed by Au deposition and lift-off processing. The latter allows for the measurement of the GMR effect in the devices (see measurement scheme detailed in Appendix \cite{Supp} in the supplemental material).

The materials used are magnetically soft and exhibit a full film coercivity of 2 Oe for the Ni$_{81}$Fe$_{19}$ films and 4 Oe for the Co$_{90}$Fe$_{10}$ films. Both materials were selected for their softness. Nonetheless, Co$_{90}$Fe$_{10}$ also exhibits an increased saturation magnetization ( M$_s$ = 1334 kA/m) compared with Ni$_{81}$Fe$_{19}$ (M$_s$ = 795 kA/m). These values were measured using a BH-looper set-up \cite{BHloop} which can detect the stray field of an entire wafer. The films are polycrystalline, with an expected crystallite size of 10 nm \cite{MAA97}.  The individual crystallites of Co$_{90}$Fe$_{10}$ can be expected to exhibit a large magneto-crystalline anisotropy constant (K = 45$\cdot 10^4$ J/m$^3$ for pure Co \cite{Cul11}) compared with Ni$_{81}$Fe$_{19}$.

A scheme of the used architecture is depicted in Fig. \ref{fig1}. The structure starts with a nucleation pad to introduce domain walls in the looping wires \cite{SMA13} and finishes with a tapered tip to prevent nucleation on that end. The devices are designed with 16 loops (i.e. the complete length of the device is 31 mm from the nucleation pad to the tapered tip end). This structure can contain a maximum of 33 domain walls (i.e. 2 domain walls per loop plus 1 in the wire following the nucleation pad), this allows for the sensing of 16 360$^\circ$-turns of the applied field.

This type of structure is very large compared to the typical dimensions encountered for domain wall-based devices in the literature \cite{GS12,LOB12,LAR15,AKP16}. The large size is advantageous for the assessment of devices reliability since the domain wall needs to cover a sizeable distance. The domain wall has thus a high probability of encountering the full distribution of geometrical and structural variations induced by the fabrication. 
Furthermore, the inner loops of the devices are far away from the starting nucleation pad and the end of the sample which could otherwise provide an unwanted extra source of device variability (reduction of the nucleation field due to flux closure at the edges). 
In order to investigate the influence of variations of the cross section, four different devices widths were compared (200, 250, 300 and 350 nm).

\section{Characterization}

After processing, a subset of the devices was characterized under Scanning Electron Microscopy (SEM) and Atomic Force Microscopy (AFM) to study the topography and assess the pattern transfer of the geometrical shape.

\begin{figure}[H]
\centering
\includegraphics[scale=0.3]{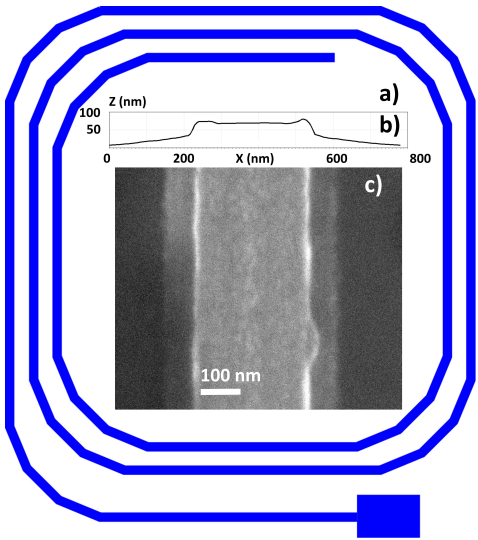}
\caption{a) Schematic of the structure, 3 loops are represented. b) Atomic Force Microscopy image of a Ni$_{81}$Fe$_{19}$ wire of 300 nm nominal width c) Scanning Electron Microscopy image of a Ni$_{81}$Fe$_{19}$ wire of 300 nm nominal width. }
\label{fig1}
\end{figure}

From the acquired images (see example in Fig. \ref{fig1}), we observe a variation of the widths compared to the nominal widths. Furthermore, the AFM profiles reveal a trapezoidal shape instead of a square shape which is a consequence of the etching process being unable to transfer precisely the photolithographically defined structures for all depths. Such variations are caused by shadowing effects during the ion milling and might affect the dynamics of the domain wall propagation as well as the nucleation process \cite{LAR15,AKP16,SG10}. Material redeposition is seen on the images as side bumps on top of the wire. The redeposition is expected to be composed of all the materials constituting the stack. It, therefore, remains difficult to determine the effective 'magnetic' width of the effective magnetic layer. In our case, we define the width as the distance between the two bright lines in the SEM image, corresponding to the bump regions of the AFM profile. The width can vary up to 60 nm between devices and a further 60 nm between the top and the bottom of the wire due to the shape of the wire profile. This effect results from the inhomogeneity of the resist thickness over the wafer and the difficulty in controlling photolithography dimensions on lengthscales below the diffraction length.

\section{Sensor operation conditions}

\subsection{MOKE investigation}

To operate the device, the processes necessary for the sensing must work flawlessly. There are two main failure events of magnetic devices, namely unwanted domain wall nucleation and domain wall pinning as a result of too low fields being applied to allow for domain wall propagation.
The Magneto-Optical Kerr Effect microscope set-up used is the following: The objective used is a x50 magnification, the source is a white linearly polarized light from an incandescent bulb source, and the microscope is operated in the longitudinal configuration. The wires of the sensor are positioned parallel to the camera's field of view to provide a reference for the angle of the applied field. A vector magnet is utilized for the application of a rotating field up to 100 mT. To detect a switching event, a differential contrast method is used. A background image is saved and subtracted to the current field of view yielding a clear contrast observable even for widths as small as 200 nm which is lower than the diffraction limit of our light source.

\subsubsection{Propagation field}

The propagation field is the lowest field value, at which the domain wall freely propagates and is not pinned at any point in the whole structure. Since domain wall pinning is a highly stochastic phenomenon \cite{MYI09, JA10} a significant amount of statistics is required to reliably characterize it. In our scheme, for a single measurement to be successful, all the 33 possible domain walls must propagate along the entire device without anyone experiencing a failure event since strong pinning of one wall necessarily leads to it being annihilated with the subsequent wall when it also reaches the pinning site. At least 33 complete rotations of the field were performed, and every domain wall probed 31 mm of wire for a pinning event. In total, 10 structures were probed for a combined length of more than 10 m of magnetic materials. The data are shown by the discs in Figure \ref{fig3}.

\begin{figure}[H]
\centering
\includegraphics[scale=0.4]{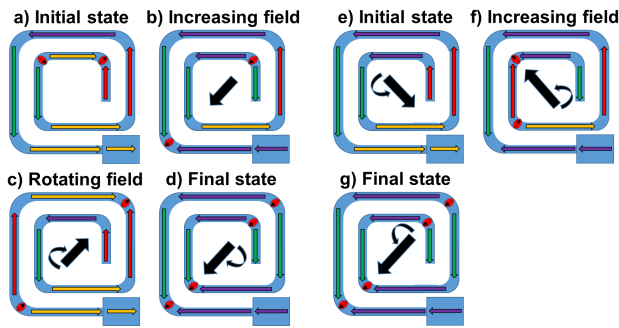} 
\caption{ Schematic representation of the field sequences for the measurement of the propagation and nucleation fields for a simplified device based on two turns. a)-d) Propagation field - a) The sensor is in the initial state with a pair of domain walls in it. The propagation field measurement can be carried out for any state of the sensor (filled with DWs or empty). b) The field is rotated clockwise and increased to inject domain walls from the nucleation pad at each 180$^\circ$ turn.  c) The nucleated domain wall should continue to propagate around the corners and further into the device as the field rotation continues. As long as the wall continues to propagate, the field is kept constant, however, if pinning is detected the field is increased to sustain the propagation. d) The device is entirely filled up with domain walls at all alternate corners in the structure. e)-g) Nucleation field measurement. e) Empty state for the initial configuration is obtained by rotating the field counter-clockwise with a field higher than the propagation field value. f) The field is increased and rotated counter-clockwise until a nucleation event occurs. g) The final state with a filled sensor.  }
\label{fig2} 
\end{figure}

All the propagation field values for the Ni$_{81}$Fe$_{19}$ samples have been found to be between 5 and 25 mT. Despite the variations in width and thickness, all the Ni$_{81}$Fe$_{19}$ samples, exhibit similar propagation fields. The shape of the wire is thus not entirely governing the domain wall propagation field values. The propagation field is mainly affected by the irregularities of the shape and the material. However, due to the redeposition on the wires, it is difficult to directly ascertain the relevant magnetic roughness of the wires, which might be different from the topographical roughness.  A priori, this is not very surprising as for perfect wires the propagation field would be zero and independent of the wire geometry. However, in real wires, defects and edge roughness play the role of governing mechanisms for the propagation field, and these effects are not strongly geometry dependent (width, thickness). The small increase with decreasing wire width can be explained by the edge roughness becoming relatively more important for narrower wires (the edge roughness is largely wire width independent).
The Co$_{90}$Fe$_{10}$ samples have propagation fields that are two times higher than the ones for the Ni$_{81}$Fe$_{19}$. We attribute the latter effect to the magneto-crystalline anisotropy of each Co$_{90}$Fe$_{10}$ crystallite generating an energy landscape which increases the pinning of the domain walls. 
Furthermore, samples with a thin layer of Co$_{90}$Fe$_{10}$ (1 nm) below the Ni$_{81}$Fe$_{19}$ layer were investigated, and we did not observe significant differences between the samples with and without the thin layer, which thus plays only a minor role for the magnetic properties.

In summary, for devices, it is therefore desirable to avoid materials with strong magneto-crystalline anisotropy and limit processing variability to ensure the reliability of domain wall propagation. The propagation field appears then as a characteristic materials parameter that is not strongly dependent on the wire width and thus cannot be easily tailored by the geometry to improve the operational reliability.
Since the propagation field does not provide for an easy handle for the improvement of magnetic sensors operation conditions as the minimum propagation field cannot be easily reduced by for instance changing the geometry, we next investigate the nucleation field of the devices.

\begin{figure}[H]
\centering
\includegraphics[scale=0.4]{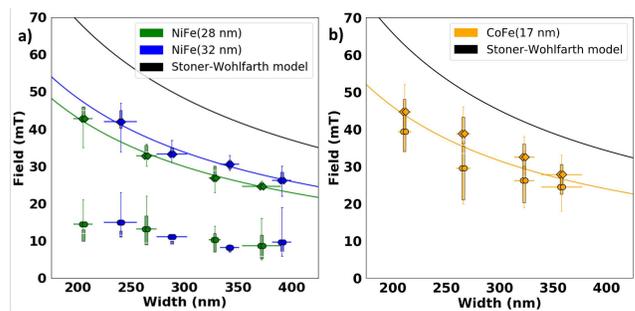} 
\caption{ Propagation and nucleation fields for the investigated samples. The boxes associated with the data points in the plot represent 25\% (first quartile) to 75 \% (third quartile) of the distribution. The whiskers or dashed lines represent 5\% to 25\% and 75\% to 95\%. In this manner, the plot represents the key features of the entire distribution. The round points represent the average value of the propagation while the diamonds represent the nucleation field. a) Plot of the nucleation and propagation field values as a function of the width of the wire for the Ni$_{81}$Fe$_{19}$(28 nm) and Ni$_{81}$Fe$_{19}$(32 nm). The black line is the pure Stoner-Wohlfarth behaviour for Ni$_{81}$Fe$_{19}$(32 nm). The blue and green lines are the adapted model fitted with a scaling constant C = 0.7. b) Similar plot as a) for Co$_{90}$Fe$_{10}$ (17 nm).}
\label{fig3}
\end{figure}

\subsubsection{Nucleation field}

At high fields, instead of domain wall nucleation only occurring in the pad region of the device, undesired domain wall nucleation takes place in the wires. For the tapered end of the device, it is expected that the shape anisotropy, in this part, is too high for the nucleation of a new domain wall for the probed field range. Having established the propagation field, we can rotate the applied field counter-clockwise to empty the device completely of all the domain walls by annihilating them in the nucleation pad.  The resulting magnetic state is the initial state, which we term the vortex state due to its resemblance to the one for ring structures as seen in Fig. \ref{fig2} e). By continuing to rotate the field in this direction with incrementing field strength, we selectively detect domain wall generation through spontaneous nucleation somewhere within the wire, as depicted in Fig. \ref{fig2} f). The field at which such an event is detected is termed the nucleation field and is of interest since it yields the field at which the measured information can potentially be lost in a failure scenario.

We find that the scaling of the nucleation field follows a power law as a function of the width. 
The nucleation field for domain walls in nanopatterned soft magnetic materials is mainly determined by the cross-sectional shape of the system (width and thickness). For a closed system such as a ring or a loop, the only boundaries are the two side edges thus which can be approximated as infinitely long.  Furthermore, if the radius of curvature is much larger than the width of the wire, then no lowering of the nucleation field value is expected at corners. Such effects are usually provoked by a flux closure spin configuration at the ends of the wire. This reduction of the nucleation field has been observed in the case where a wire relaxes into an S or C state \cite{Had12}. 
Since our materials are relatively thick and soft, the magnetization is lying in the plane in the direction of the wire length due to a dominant magnetostatic energy contribution. The expectation is that the shape anisotropy is playing a primary role in the determination of the nucleation field value. 
Within the framework of the Stoner-Wohlfarth model, a particle with dimensions smaller than the exchange length (5 nm in Ni$_{81}$Fe$_{19}$), the magnetization is approximated with a macrospin and is expected to rotate coherently during the switching. In the most simplistic version of the model, this particle is only subject for instance to a uniaxial anisotropy of the form $K\cdot sin^2(\theta)$, in our case, K being mainly the shape anisotropy. 

For larger systems that are not fully described as a macrospin, one can expect an activation volume, located for instance at the point of lowest anisotropy, to rotate coherently. Mathematically, this can be described as follows:

\begin{equation}\label{eq:2}
E = E_{Zeeman} + E_{Demag} = -\mu_0 H M_s V + \frac{1}{2} \mu_0 N_y M^2_s V 
\end{equation}

with $M_s$ being the saturation magnetization, $V$ the activation volume and $N_y$ the demagnetizing factor described in \cite{JAO45} for an infinitely long wire:

\begin{equation}\label{eq:3}
H_n = \frac{1}{2} \frac{t}{t+w} M_s
\end{equation}

\begin{figure}[H]
\centering
\includegraphics[scale=0.4]{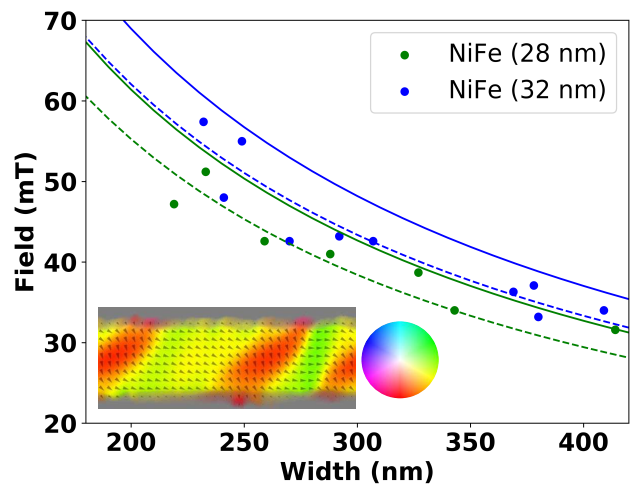} 
\caption{Simulation results of the nucleation field as a function of the minimum width of the wire. The green and blue full line represent the SW model for the Ni$_{81}$Fe$_{19}$(28 nm) and Ni$_{81}$Fe$_{19}$(32 nm), respectively. The dashed lines are the SW model with a scaling factor C = 0.9. A simulation snapshot of the magnetization was taken at a field just 1 mT below the nucleation field value of a 200 nm wide wire of Ni$_{81}$Fe$_{19}$(32 nm).}
\label{fig4}
\end{figure}

With $t$ the thickness and $w$ the width of the sample. In this simple formula, the nucleation field is then determined by the geometry as well as the saturation magnetization, which is a materials constant. As plotted, in Fig. \ref{fig3} in black and Fig. \ref{fig4} full lines, such a calculated curve for the pure Stoner-Wohlfarth behaviour does not reproduce the obtained data quantitatively. 
 
To understand the magnetization behavior close but below the nucleation field, some simulations are performed with the software Mumax3 \cite{Van14}. AFM profiles are used as the simulated shape, and periodic boundary condition serves to extend the length of the wire. The framework is discretized with a cell volume of 2x2x2 nm$^3$ to permit a good representation of the realistic shape. 
The magnetization is initialized upward and left to relax. A field is then applied for 20 ns following an equation of the form $B = (-\frac{\sqrt{2}}{2} B_{ext}(1-\exp(\frac{-t}{4e^{-9}})),-\frac{\sqrt{2}}{2} B_{ext}(1-\exp(\frac{-t}{4e^{-9}})), 0)$ to avoid artefacts due to an instantaneous applied field. A bisection method was then used to determine the nucleation field within 1 mT precision.

The results are presented in Fig. \ref{fig4}. A snapshot of the magnetization in Fig. \ref{fig4} represents the relaxed state of the stripe at an applied field value that is just 1 mT below the nucleation field value. The spin structure and the subsquent dynamics show that the reversal mode of our stripes ressembles an in-plane curling mode. The latter is expected for large systems with inhomogeneities in the anisotropies \cite{Wer01}. As compared to the Stoner-Wohlfarth model, the rotation of magnetization is not coherent in the whole structure and exhibits a 2-dimensional variation along the wire \cite{Kro87}. A pure curling mode would not yield nucleation field values significantly different from the Stoner-Wohlfarth model \cite{Aha99}. Furthermore, our data also show that despite the trapezoidal shape and the included edge roughness, the nucleation field at 0 K is on average 90 $\%$ of the expected one from the Stoner-Wohlfarth model.

However, wire irregularities are expected to yield a lowered nucleation field as seen from the inhomogeneous spin configurations shown in the simulation results in Fig. \ref{fig4}. In the real system, further effects can lead to a reduction of the nucleation field. Damage from the ion milling causing a change of crystallization at the edges and a decrease of the saturation magnetization or a doping of the wire due to material implantation can lead to a locally reduced shape anisotropy due to a reduced saturation magnetization. Finally, thermal activation is also going to have an impact not considered in the Stoner-Wohlfarth model. In order to account for such effects in the simplest manner, we include a scaling factor C to the demagnetization factor as follows:

\begin{equation}\label{eq:4}
H_n = C \frac{1}{2} \frac{t}{t+w} M_s
\end{equation}

By fitting the data in Fig. \ref{fig3}, we find that a value of C = 0.7 provides good agreement with the results of the samples. Overall, the clearly observable variations of the nucleation field with the geometry provide a handle to tailor the characteristics of the device thus the operation condition and the reliability.

\subsection{GMR investigation}

To investigate this further and obtain better statistics, we measure a very large number (3200) of 16 loop devices. The majority of the possible geometrical variations are going to be encountered, generating a clear idea of the maximum deviations still enabling a working device.
Furthermore, by taking the absolute resistance value of the sensor based on the geometry (width, thickness, and length), we can compare it to the nucleation field to check for possible connections. 

An entire wafer was prepared with the previously mentioned GMR stack structures and contacted to measure the resistance of the wires. The resistance in the initial state (no domain walls) is probed and compared to the case where a domain wall is inside. An algorithm using also a bisection method is used to define the exact nucleation field value. The applied field sequence is made of 17 rotations counter-clockwise followed by an electrical measurement. After the 17 turns, the device is expected to be empty if the nucleation field is higher than the tested value. We describe the standard measurement scheme for a working element, any failure in the steps result in the device being counted as defective. A starting value of 30 mT is used since the propagation limit was previously measured lower than this. After the sequence, if the sensor is measured empty then the field is increased to 80 mT, and the sensor should be entirely filled with domain walls as we expect the nucleation field of all the structures to be lower. The next field value is taken as half of the difference of the previous ones added to the lowest value giving 55 mT. If the device is measured empty, then the following field is half of the difference between the highest (80 mT) and the middle value (55 mT) added to the middle value. The lowest value (30 mT) is replaced by the middle one (55 mT). If the sensor is filled, then half of the difference between the lowest (30 mT) and the middle value (55 mT) is added to the lowest value. The highest value (80 mT) is replaced by the middle one (55 mT). The algorithm continues until the difference between the lowest and the highest field values is smaller than 1 mT.

The latter process served to measure the specific nucleation field of 3200 structures. 
Approximately 800 measurements per wire width were performed. 
The resistance is measured across the device between the V$_{cc}$ and GND (as seen in the Appendix \cite{Supp}). Due to the connection layout, the number of wire connected is 33, and their length is 400 $\upmu$m. Before the resistance measurement, the sensors were initialized with domain walls in the whole device. 
Due to the looping configuration and the domain walls present, half of a wire is in a ‘high’ resistive state with the GMR and the other half is in a ‘low’ resistive state. Thus the measured resistance does not contain a GMR component (the DW has a negligible contribution). The resistance of every wire is then similar. 
We then apply the formula $r = R_s  \frac{L}{w}$ for the resistance of a wire, with  R$_s$ = 4.09 $\Omega$/sq being the sheet resistance, L the length of a wire and w the width. 
The wires are connected in parallel thus the resistance of the device is $R = \frac{r}{33}$. We plug the latter in equation \ref{eq:4} to obtain:

\begin{equation}\label{eq:5}
H_n = \frac{1}{2} \frac{t}{R_sL+33Rt} 33M_sR
\end{equation}

with $R$ being the resistance of the sensor, $t$ the thickness, $M_s$ the saturation magnetization and $L$ the length of the wire.
The previously described model with a constant $C$ = 0.7 is plotted in brown while the Stoner-Wohlfarth nucleation field is in black.

\begin{figure}[H]
\centering
\includegraphics[scale=0.4]{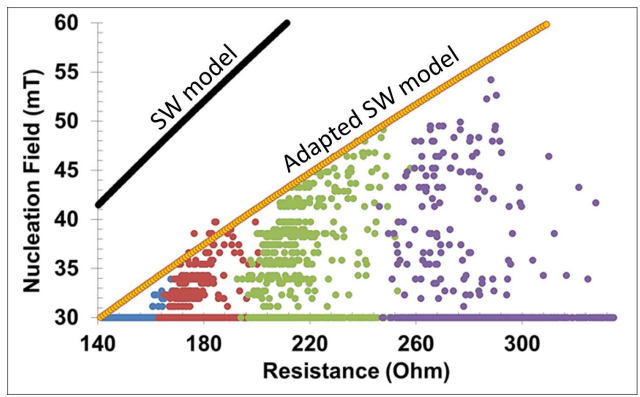}
\caption{
Nucleation field of the device as a function of the resistance. The different nominal widths are represented by different colors: 350 nm (blue), 300 nm (red), 250 nm (green) and 200 nm (purple). 
}
 \label{fig5}
\end{figure}

In the Fig. \ref{fig5}, a fixed reference value for propagation is used (30 mT). Indeed, for the algorithm to work a starting propagation field value is needed. 

The expected theoretical resistances are 247 $\Omega$, 198 $\Omega$, 165 $\Omega$, 141 $\Omega$ for respectively 200 nm (purple), 250 nm (red), 300 nm (green), 350 nm (blue). The measurement of the resistance is realized over the whole structure with the 33 wires in a parallel configuration. The resistance of the device is then plotted as a function of the nucleation field. The previously described model is shown as well as the Stoner-Wohlfarth model. 

Only a few points are visible in the figure for the 350 nm wire width due to an average nucleation lower than 30 mT. The nucleation field values are observed to lie below the adapted model. 
Furthermore, we observe that the resistance is always higher than the one expected from the nominal width. The latter is a confirmation that the measured width with the SEM is probably not the effective electrical width contributing to the magnetic and electrical signal and that most of the sides are covered with redeposited material thus generating an effectively larger topographical width. Note that taking into account the multilayer nature of the GMR stack by calculating the resistance using the Fuchs-Sondheimer model \cite{KF38, EHS52} does not significantly change the values as the conduction is dominated by the thick Ni$_{81}$Fe$_{19}$ free layer which carries most of the current.
The limit shown by the adapted Stoner-Wohlfarth model demonstrate the impossibility for our current architecture (3 vertices and 16 loops of the geometries used) to reach the ideal Stoner-Wohlfarth model behaviour. The fitting constant of 0.7 sets the maximum average nucleation field obtainable for these industrially produced devices in these geometries. These results are of major importance for applications due to the fact that a simple resistance measurement allows to the identification of a non-working device. As an example, if the requirements are that devices should not exhibit a nucleation field lower than 40 mT then any device with a resistance below 200 $\Omega$ can be discarded. This provides a tremendous gain of time since the measurement duration for the characterization of a single device can be 1 minute.

\section{Conclusion}

To conclude, we have determined the critical fields for sensor operation based on domain wall propagation allowing us to gauge the limitations for the operation of the devices. We find that both the materials and the geometry play a key role. Firstly, the origin of the geometrical dependence of the propagation is difficult to pinpoint, since a variety of factors contribute to the variation in propagation field and these are hard to precisely characterize and quantify. Compared to Ni$_{81}$Fe$_{19}$, the Co$_{90}$Fe$_{10}$ samples yield a drastically increased propagation field. This can be attributed to the enhanced magneto-crystalline anisotropy in each individual crystallite compared to Ni$_{81}$Fe$_{19}$.

For the nucleation field, the dependence on the geometry exhibits a geometrical scaling of the form that one expects if the shape anisotropy dominates. It can be described by the Stoner-Wohlfarth model, despite the simulation not showing a coherent rotation of the magnetization in the complete stripe. Thus the nucleation field geometry dependence provides a handle for the improvement of magnetic domain wall sensors.
For all the measured materials, the maximum expected nucleation value can be fitted by a corrected uniaxial anisotropy rectified by a constant accounting for the processing, the angle segmentation, and the geometrical scale. 
A measurement of a large number of devices is performed to allow for an accurate assessment of the fitting constant used for the MOKE measurement results. We can ascertain a definite limit by the previously mentioned maximum nucleation field value and find that this is related to the resistance of the sensor.  This limit can be used as a tool for future fast analysis of magnetic sensors.

\section*{Acknowledgements}
The authors would like to acknowledge the company Sensitec GmbH for providing the samples, the company Novotechnik for providing the designs and the WALL project for financial support. The work and results reported in this publication were obtained with research funding from the European Community under the Seventh Framework Programme - The people Programme, Multi-ITN “WALL” Contract Number Grant agreement no.: 608031, and a European Research Council Proof-of-Concept grant (MultiRev ERC-2014-PoC (665672)).

\newpage


%

\end{document}